\documentclass[article]{revtex4}

\usepackage{enumerate}
\usepackage{enumitem}
\usepackage{graphicx}
\usepackage{dcolumn}
\usepackage{bm}
\usepackage{graphicx}
\usepackage{amssymb}
\usepackage{epstopdf}
\usepackage{color}

\usepackage[english]{babel}
\usepackage{xspace}
\newcommand{\LTO}{$\mbox{LaTiO}_3$\xspace}
\newcommand{\STO}{$\mbox{SrTiO}_3$\xspace}
\newcommand{\LAO}{$\mbox{LaAlO}_3$\xspace}
\newcommand{\TiO}{$\mbox{TiO}_2$\xspace}

\newcommand{\VG}{$V_{\mathrm{G}}$\xspace}

\begin{document}

\title{Multiple Quantum Phase Transitions at the superconducting \LTO/\STO interface}

\author{J. Biscaras$^1$, N. Bergeal$^1$, S. Hurand$^1$, C. Feuillet-Palma$^1$, A. Rastogi$^2$, R. C. Budhani$^2$$^,$$^3$, M. Grilli$^4$, S. Caprara$^4$, J. Lesueur$^1$}

\affiliation{$^1$LPEM- UMR8213/CNRS - ESPCI ParisTech - UPMC, 10 rue Vauquelin - 75005 Paris, France}
\affiliation{$^2$Condensed Matter - Low Dimensional Systems Laboratory, Department of Physics, Indian Institute of Technology Kanpur, Kanpur 208016, India}
\affiliation{$^3$National Physical Laboratory, New Delhi - 110012, India }
\affiliation{$^4$Istituto Nazionale per la Fisica della Materia, Unit\`{a} di Roma 1 and SMC Center, and Dipartimento di Fisica Universit\`{a} di Roma
``La Sapienza'', piazzale Aldo Moro 5, I-00185 Roma, Italy}

\date{\today}

\begin{abstract}
We study the magnetic field driven Quantum Phase Transition (QPT) in electrostatically gated superconducting \LTO/\STO interfaces. Through finite size scaling analysis, we show that it belongs to the (2+1)D~XY model universality class. The system can be described as a disordered array of superconducting islands coupled by a two dimensional electron gas (2DEG). Depending on the 2DEG conductance tuned by the gate voltage, the QPT is single (corresponding to the long range phase coherence in the whole array) or double (one related to local phase coherence, the other one to the array). By retrieving the coherence length critical exponent $\nu$, we show that the QPT can be ``clean'' or ``dirty'' according to the Harris criteria, depending on whether the phase coherence length is smaller or larger than the island size. The overall behaviour is well described by a theoretical approach of Spivak \emph{et al.}, in the framework of the fermionic scenario of 2D superconducting QPT. 
\end{abstract}

\maketitle

\indent

Two-dimensional electron gas at the interface between insulating oxides (O-2DEG) have raised considerable interest\cite{Mannhart:2010ha,Mannhart:2008uj,Hwang:2012nm}. Indeed, they display very high mobilities suitable for applications\cite{Ohtomo:2004hm}, but also a rich phase diagram, with different quantum ground states such as magnetism\cite{Brinkman:2007fk,Shalom:2009fi,Bert:2011cg,Li:2011jx} or superconductivity\cite{Reyren:2007gv,Biscaras:2010fp} for example. Moreover, these properties can be finely tuned using a gate voltage\cite{Caviglia:2008uh,Biscaras:2012eu}. 
Therefore, O-2DEG appear as interesting systems to study Quantum Phases Transitions (QPT) that occur between different quantum states, when a parameter in the Hamiltonian crosses over a critical value\cite{Sondhi:1997wz}. The critical behaviour of the observables belong to universality classes which depend on general properties of the system such as its dimensionality or its symmetries, and not on the microscopic details. The associated critical exponents obey specific rules, among which the so-called ``Harris criteria''. The latter stipulates that the correlation length exponent $\nu$ must satisfy $\nu\geq2/d$ for ``dirty'' disordered systems and $\nu\leq2/d$ for ``clean'' ones, where $d$ is the spatial dimensionality\cite{Harris:1974tn}. 

\indent An important example of QPT is the transition from a superconducting to an insulating state in two dimensions, which has been a matter of debate for a long time\cite{Goldman:2010gc}. Numerous experiments with contrasted results\cite{Goldman:2010gc} have been performed, and a large variety of critical exponents have been found. The nature of the non-superconducting state and the role of the disorder are still unclear. The possibility of observing multiple QPT has been raised recently\cite{Spivak:2001bk,Spivak:2008cc,Feigelman:472483}. Here we show that a perpendicular magnetic field applied on a superconducting O-2DEG drives the system towards a weakly localized metal. We evidence two QPT corresponding to the ``dirty'' and ``clean'' limits of the Harris criteria, which can be controlled by the gate voltage. 
More precisely, we have shown that the QPT in superconducting O-2DEG belongs to the (2+1)D~XY universality class\cite{Sondhi:1997wz}, and follows a fermionic scenario\cite{Finkelshtein:1987vo,Yazdani:1995ut}. Finite size scaling analysis (FSS) reveals that depending on the gate voltage and the temperature, the product of the critical exponents $z\nu$ ($z$ is the dynamical exponent) is $z\nu=2/3$ in the clean regime in agreement with calculations\cite{LI:1989uu}, and greater than one otherwise, as expected from the Harris criteria. We argue that this multiple QPT can be understood in the frame of a recent model developed by Spivak \emph{et al.}\cite{Spivak:2008cc} to describe the QPT in a disordered array of superconducting puddles coupled by a 2DEG.This result demonstrates that the superconducting O-2DEG tunability is a powerful tool to study fundamental properties of superconductors in low dimensions. 
We anticipate that further insight within QPT involving two dimensional superconductors will emerge. For example, the role of disorder and inhomogeneity\cite{Ghosal:1998up,Spivak:2008cc}, Coulomb repulsion, charge fluctuations and screening on the QPT\cite{Goldman:1998jy,Kapitulnik:2001vd,Vishwanath:2004ed} can be explored in more details.\\

\begin{figure}[h]
\includegraphics[width=12cm]{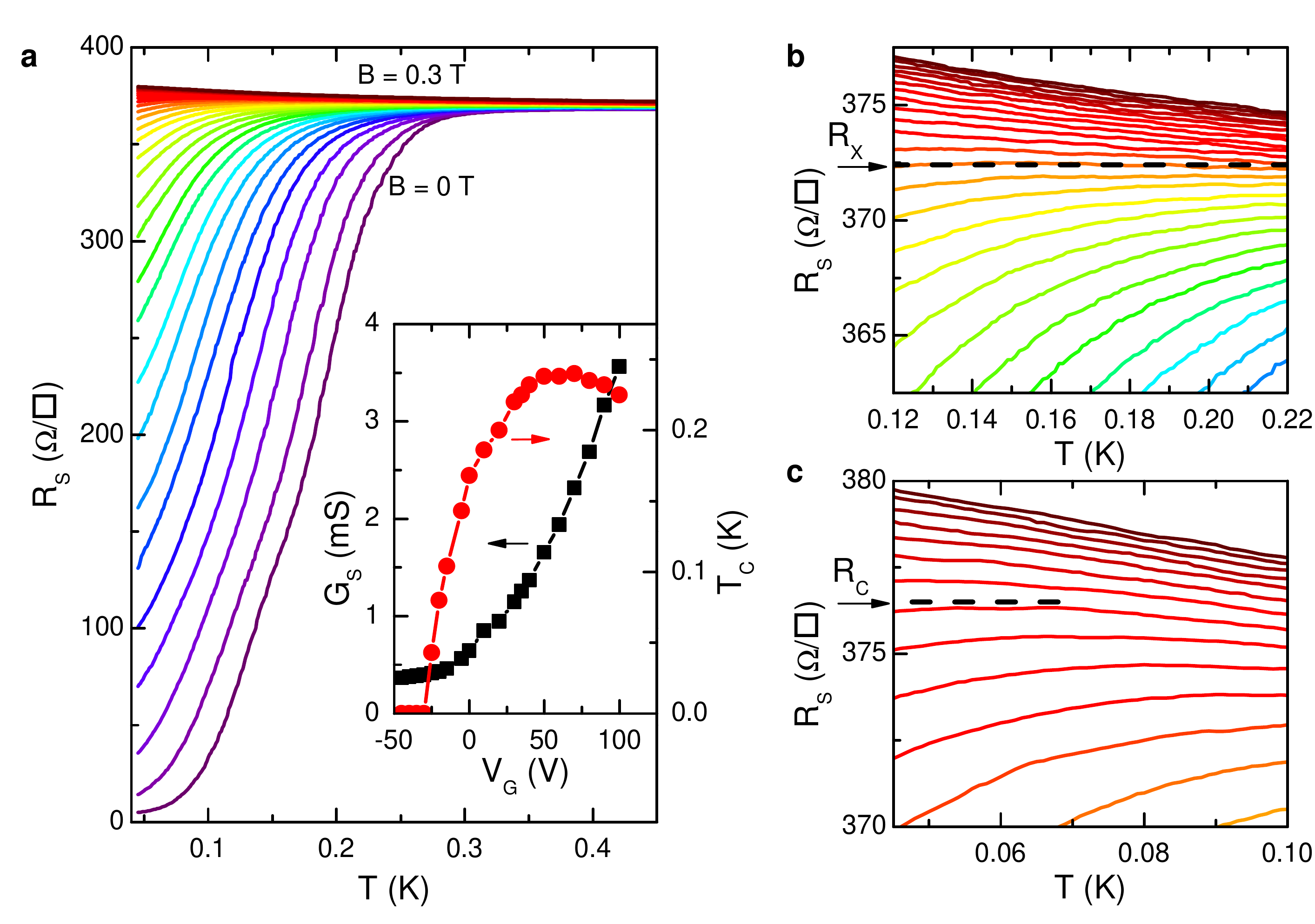}
\caption{{\bf Superconductor-insulator transition induced by a magnetic field.} {\bf a,} Sheet resistance $R_{S}$ as a function of temperature for different magnetic fields from 0 to 0.3 T. {\bf Inset :} $T_c$ and $G_S$=1/$R_S$ as a function of the gate voltage \VG. {\bf b,} Zoom on the same data showing the characteristic magnetic field $B_{\times}$, for which $R_{S}$ is constant between 0.12~K and 0.22~K. {\bf c,} Zoom on the same data showing the characteristic magnetic field $B_{\mathrm{C}}$, which separates the two regimes at the lowest temperatures.}
\label{Figure1}
\end{figure}

\begin{figure}[h]
\includegraphics[width=12cm]{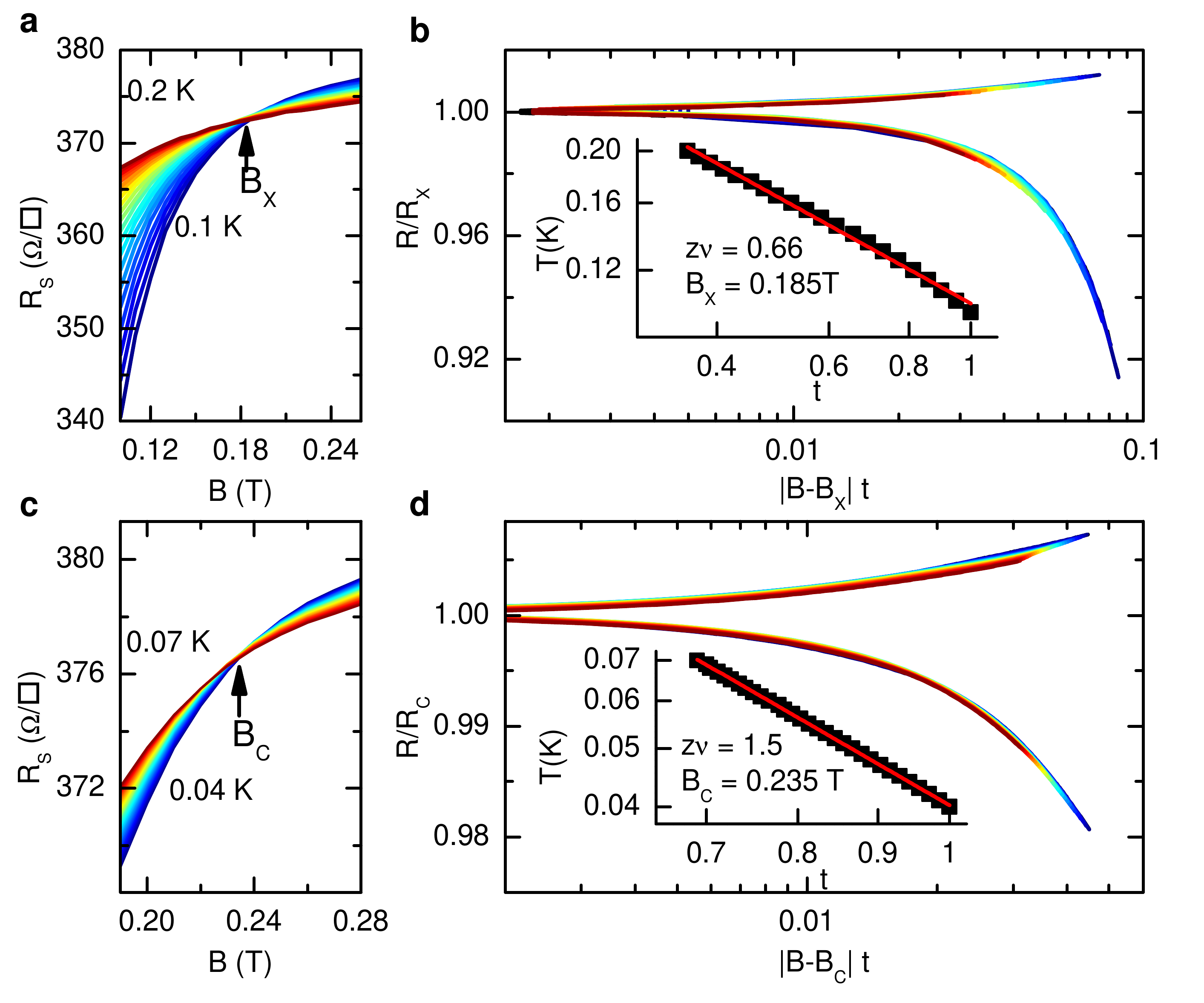}
\caption{{\bf Finite size scaling analysis for \VG=+80~V.} {\bf a,} Sheet resistance $R_{s}$ as a function of magnetic field $B$ for different temperatures from 0.1 to 0.2~K. The crossing point is ($B_{\times}$=0.185~T, $R_{\times}$=372.4~$\Omega/\square$).{\bf b,} Finite size scaling plot $R_{S}/R_{\times}$ as a function of ${\vert B-B_{\times} \vert t}$ (see text for the definition of $t$). {\bf Inset :} temperature behaviour of the scaling parameter $t$. The power law fit gives $z\nu$=0.66. {\bf c,} Sheet resistance $R_{S}$ as a function of magnetic field $B$ for different temperatures from 0.04 to 0.07~K (\VG=+80~V). The crossing point is ($B_{\mathrm{C}}$=0.235~T, $R_{\mathrm{C}}$=376.6~$\Omega/\square$). {\bf d,} Finite size scaling plot $R_{S}/R_{\mathrm{C}}$ as a function of ${\vert B-B_{\mathrm{C}} \vert t}$. {\bf Inset :} temperature behaviour of the scaling parameter $t$. The power law fit gives $z\nu$=1.5}
\label{Figure2}
\end{figure}

\indent 
The \LTO/\STO epitaxial interface exhibits a high mobility O-2DEG which displays superconductivity\cite{Biscaras:2010fp}. 
We recently showed that $T_c$ can be tuned by electrostatic gating from its maximum value of $\sim$200~mK to zero, and that superconductivity coincides with the presence of Highly Mobile Carriers (HMC) at the edge of the quantum well formed by the O-2DEG at the interface\cite{Biscaras:2012eu}.
 It is therefore possible to prepare the system with a given $T_c$ by controlling the gate voltage \VG (inset Fig.~\ref{Figure1}a), and to study how the superconducting state is destroyed by a perpendicular magnetic field in this situation.
Samples are grown by pulsed laser deposition of 15 unit cells of \LTO on \TiO-terminated (001)\STO substrate (see Biscaras \emph{et al.}\cite{Biscaras:2010fp} for details). A metallic back gate is evaporated at the rear of the 500~$\mu$m thick \STO substrate and connected to a voltage source (\VG).
Standard four-probe resistance measurements are made with a current sufficiently low to avoid any heating of the electron at the lowest temperature.
The polarization scheme described previously\cite{Biscaras:2012vy} is applied to insure a reversible behaviour of the superconducting O-2DEG.
The normal state displays the logarithmic temperature dependence of the conductivity, characteristic of weak localization in two dimensions\cite{Biscaras:2010fp}.
Hence, the magnetic field turns the 2D superconductor into a 2D weakly localizing metal as shown in Fig.~\ref{Figure1}a for \VG=+80~V, where the resistance per square $R_{S}$ is plotted as a function of temperature $T$ for different perpendicular magnetic fields $B$.
A close-up view of the data reveals a critical field $B_{\times}$ which separates the two regimes, and for which $R_S$ is constant (Fig.~\ref{Figure1}b).
The same set of data plotted as $R_{S}$ versus $B$ for different temperatures in Fig.~\ref{Figure2}a displays a crossing point ($R_{\times}$=372.4~$\Omega/\square$, $B_{\times}$=0.185~T) where the resistivity does not depend on temperature. This is a first signature of a continuous QPT.

\indent In such zero temperature transitions, the ground state of the Hamiltonian is changed by an external parameter (for now magnetic field). Close to the transition the correlation length $\xi$ in the space dimensions and the dynamical correlation length $\xi_{\tau}$ in the imaginary time dimension of the quantum fluctuations diverge with a power law dependence of the distance from the transition $\delta = \vert B - B_{\times}\vert $\cite{Sondhi:1997wz,Goldman:2010gc}. At $T$=0 the correlation length exponent $\nu$ defined as $\xi \propto \vert \delta \vert ^ {-\nu} $ and the dynamical scaling exponent $z$ defined as $\xi_{\tau}\propto \xi^z$ are believed to be independent of the microscopic details of the transition and only depend on few properties of the system, such as the dimensionality and the range of the interactions, which define universality classes for the QPT\cite{Sondhi:1997wz}. The effective dimensionality of the system is $d+z$ where $d$ is the spatial dimensionality. At finite temperature, the imaginary time dimension is limited by the temperature fluctuations so that the dimensionality of the system is $d+z$ at $T$=0 only, and $d$ at finite temperature\cite{Goldman:2010gc}. More precisely, the finite temperature limits the size of the temporal direction by the thermal cut-off $L_{\tau} = \hbar/k_{\mathrm{B}} T$ which is now an upper bound for the dynamical correlation length $\xi_{\tau}$ near the critical point. It follows that in the spatial dimensions the quantum fluctuations lose phase coherence over a temperature dependent dephasing length $L_{\phi} \propto 1/T^{1/z}$ \cite{Sondhi:1997wz}.  This leads to a so-called finite size scaling (FSS) of the observables of the system. For instance, the resistance takes the form :
\begin{equation}
\frac{R_S}{R_{\times}} = F\left(\frac{\vert B-B_{\times} \vert}{T^{1/z\nu}}\right)
\label{eq1}
\end{equation}
where $F$ is a universal function with $F$(0)=1 \cite{Fisher:1990vl}. The critical exponents can then be retrieved by a scaling procedure\cite{Goldman:2010gc} as follows. The resistance is rewritten as $R(\delta,t)= R_{\times} F(\vert \delta \vert t)$ with $t$ an unknown parameter that depends only on $T$. The parameter $t$ is then found at each temperature $T$ by optimizing the collapse around the critical point between the curve $R(\delta,t(T))$ at temperature $T$ and the curve $R(\delta,t(T_0))$ at the lowest temperature considered $T_0$, with $t(T_0)$=1. The dependence of $t$ with temperature should be a power law of the form $t= (T/T_0)^{-1/z\nu}$ in order to have a physical sense, thus giving the critical exponent product $z\nu$. The interest of this procedure is to perform the scaling without knowing the critical exponent beforehand. The result of this procedure applied to the data displayed in Fig.~\ref{Figure1}a shows that data collapse onto a single (bi-valued) curve in Fig.~\ref{Figure2}b, and yields $z\nu$=0.66 (inset Fig.~\ref{Figure2}b).\\

In the literature, $z\nu$=2/3 has been observed for perpendicular magnetic field driven transitions in conventional 2D disordered superconductors such as a-NbSi\cite{Aubin:2006wk,MarracheKikuchi:2008cka} or a-Bismuth\cite{Markovic:1998kz,Goldman:2010gc}. This value corresponds to the ``clean'' (2+1)D~XY universality class according to numerical simulations\cite{Kisker:1997gt,CHA:1994vf,LI:1989uu}, where the extra dimension refers to the imaginary time in the quantum transition\cite{Sondhi:1997wz}. In that case, long range superconducting correlations are destroyed by quantum phase fluctuations. In general, the dynamical exponent $z$ is found to be one, corresponding to long range Coulomb interaction between charges\cite{Fisher:1990vl,Sondhi:1997wz}, as it has been measured in a-MoGe for instance\cite{Yazdani:1995ut}. In the absence of specific screening or dissipation mechanism\cite{Mason:2002fg}, there is no reason to invoke short range interactions which would set $z\not=1$\cite{Sorensen:1992gu}. Therefore, if $z=1$, then $\nu=2/3$, which means that $\nu \leq 2/d$, with the spatial dimension $d=2$. According to the very general Harris criteria\cite{Harris:1974tn}, the system is in the ``clean'' limit at the relevant scale for the transition, namely the dephasing length $L_{\Phi}$. In 2D disordered systems, i.e. in the ``dirty'' regime, $\nu$ is expected to increase beyond 1\cite{Kisker:1997gt,CHA:1994vf}, following the Harris criteria, as observed in a-MoGe\cite{Yazdani:1995ut}, InOx\cite{Steiner:2008ke} or ultra-thin High Tc superconductors\cite{Bollinger:2011eq} for instance.

\indent We now focus on the same $R_{S}$ vs $T$ curves below 0.1~K (\VG=+80~V). A blow-up in Fig.~\ref{Figure1}c evidences another critical field $B_{\mathrm{C}}$ for which the resistance is constant in a restricted range of temperature and $\frac{\partial R}{\partial T}$ changes sign. Reported on Fig.~\ref{Figure2}c, $R_{S}$ vs $B$ curves display a crossing point ($B_{\mathrm{C}}$=0.235~T, $R_{\mathrm{C}}$=376.6~$\Omega/\square$), and the FSS analysis shows a good collapse of the curves, leading to a critical exponent product $z\nu$=1.5 (Fig.~\ref{Figure2}d). This is clearly greater than one. The system is in the dirty limit of the Harris criteria at the scale $L_{\Phi}$.
As a conclusion for \VG=+80~V, the system undergoes a ``clean'' QPT ($z\nu$=2/3) whose critical behaviour can be seen between 120~mK and 220~mK, and a ``dirty'' QPT ($z\nu$=3/2) at lower temperature.\\

\begin{figure}[h]
\includegraphics[width=12cm]{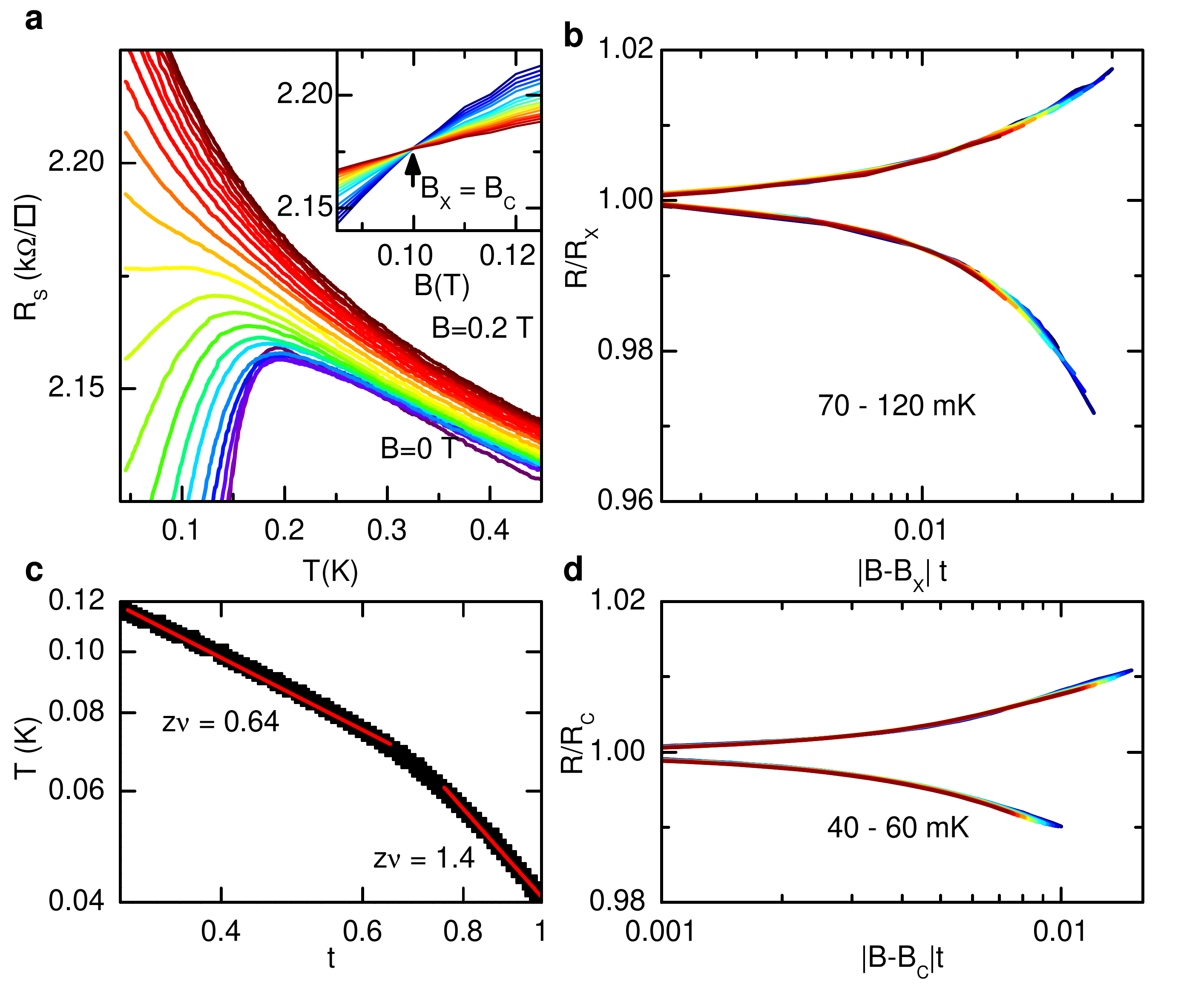}
    \caption{{\bf Finite size scaling for \VG=-15~V.} {\bf a,} Sheet resistance $R_{S}$ as a function of temperature for different magnetic fields from 0 to 0.2~T. {\bf Inset :} corresponding $R_{S}$ as a function of the magnetic field for different temperatures from 0.07~K to 0.12~K. The crossing point is ($B_{\times}$=0.1~T, $R_{\times}$=2176~$\Omega/\square$) . {\bf b,} Temperature behaviour of the scaling parameter $t$ : two distinct slopes are evidenced, 0.64 at high temperature and 1.4 at low temperature. {\bf c,} Finite size scaling plot $R_{S}/R_{\times}$ as a function of ${\vert B-B_{\times} \vert t}$ corresponding to the ``high temperature'' regime, with $z\nu$=0.64 . {\bf d,} Finite size scaling plot $R_{S}/R_{\mathrm{C}}$ as a function of ${\vert B-B_{\mathrm{C}} \vert t}$ corresponding to the ``low temperature'' regime, with $z\nu$=1.4}
\label{Figure3}
 \end{figure}

When the gate voltage is tuned to lower or even negative values, the system is driven toward a more resistive state, and $T_c$ decreases (see inset Fig.~\ref{Figure1}a). 
In Fig.~\ref{Figure3}a, $R_{S}$ is displayed as a function of the temperature for different magnetic fields and for a gate voltage \VG=-15~V. A plateau is clearly seen down to the lowest temperature for a critical field $B_{\times}$=0.1~T ($R_{\times}$=2176~$\Omega/\square$), as confirmed by the presence of a crossing point (inset of Fig.~\ref{Figure3}a). The FSS analysis reveals that two regimes take place as shown in Fig.~\ref{Figure3}b. Indeed, two distinct $z\nu$ values can be extracted : at ``high'' temperature (70~mK to 120~mK) $z\nu$=2/3 and at the lowest temperatures $z\nu$=3/2. The corresponding collapses of the data depicted in Fig.~\ref{Figure3}c,d confirm the existence of the QPT.

\begin{figure}[h]
\includegraphics[width=12cm]{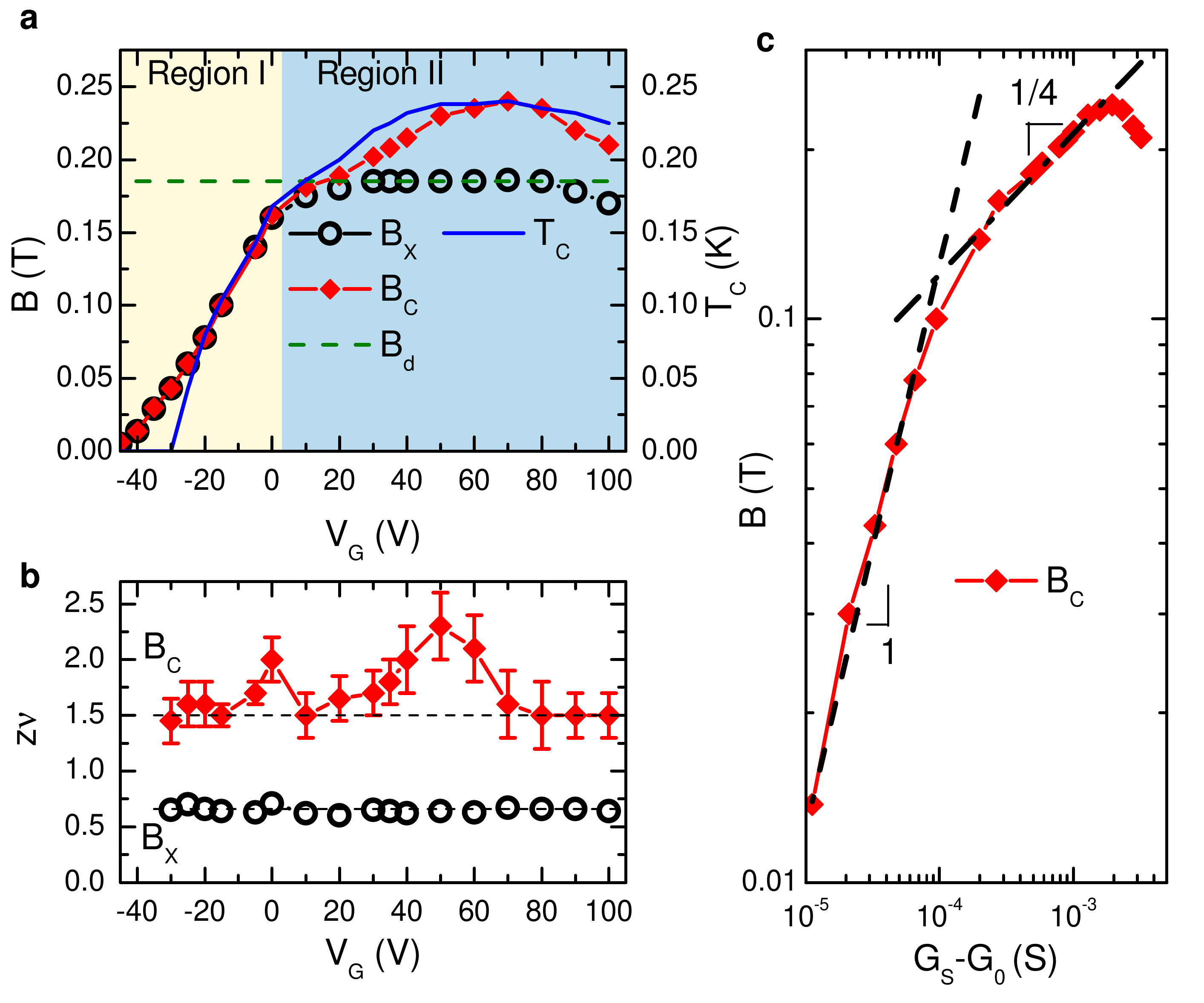}
    \caption{{\bf The two quantum phase transitions.} {\bf a,} $B_{\times}$, $B_{\mathrm{C}}$ (left scale) and $T_c$ (right scale) as a function of \VG. The dashed line corresponds to $B_d$ (see text). Regions I and II refers to the low, and respectively high, coupling regimes (see text). {\bf b,} $z\nu$ as a function of \VG for the two transitions (when $B_{\times}$=$B_{\mathrm{C}}$, $B_{\mathrm{C}}$ has been assigned to the ``low temperature'' regime, and $B_{\times}$ to the ``high temperature'' one). $z\nu$ is mainly constant with $z\nu$=0.66 for the ``high temperature'' transition. It is between 1.5 and 2.3 for the dirty one. The dashed lines correspond to $z\nu$=2/3 and $z\nu$=3/2. {\bf c,} $B_{\mathrm{C}}$ as a function of $G_S$ on a log scale. Dashed lines have slopes 1 and 1/4 respectively as expected from the SOK model.}
\label{Figure4}
 \end{figure}

\indent We made the same analysis for all the gate voltages between \VG=-40~V and \VG=+100~V. The results are presented in Fig.~\ref{Figure4}. For \VG$\geq$+10~V, two distinct critical fields $B_{\times}$ and $B_{\mathrm{C}}$ ($B_{\mathrm{C}} > B_{\times}$) can be found (region II in Fig.~\ref{Figure4}a), whereas for \VG$\leq$+10~V, they merge into a unique value (region I). 
Two important observations need to be made : 
(i) $B_{\times}$ increases with \VG and then saturates when $B_{\times}\not= B_{\mathrm{C}}$ to a value $B_d$.
(ii) $B_{\mathrm{C}}$ matches $T_c$ defined as the temperature where the normal state resistance drops of 10\%\cite{Biscaras:2012eu}. $B_{\mathrm{C}}$ is therefore the critical field which fully destroys superconductivity in the system.The critical exponent product $z\nu$ extracted from the FSS around both critical fields are reported as a function of \VG in Fig.~\ref{Figure4}b. $z\nu$ corresponding to $B_{\times}$ is constant and equals to 2/3 : the QPT is in the clean limit. For $B_{\mathrm{C}}$, $z\nu$ changes with \VG, with rather important error bars due to the finite scaling range in temperature, but is \emph{always larger than 1}, indicating that the QPT is in the dirty limit.\\

\begin{figure}[h]
\includegraphics[width=10cm]{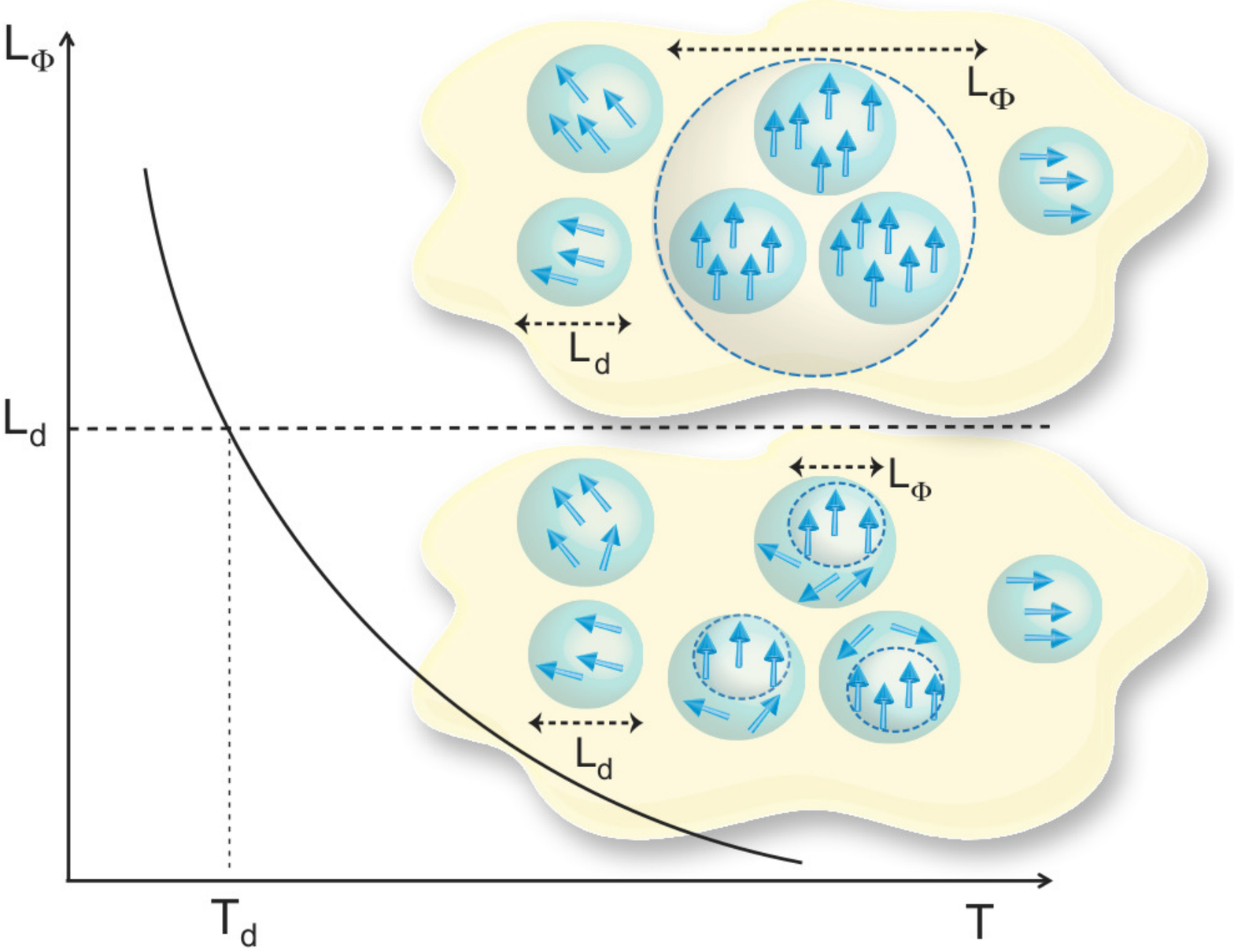}
    \caption{{\bf Sketch of the role of mesoscopic disorder on the QPT.} The dephasing length $L_{\Phi}$ diverges with temperature, and reaches the size of the superconducting puddles $L_d$ at $T_d$. The insets show a piece of material in the two regimes. At high temperature (bottom), $L_{\Phi}<L_d$ and the system is in the clean limit, whereas at low temperature (top), $L_{\Phi}>L_d$, and the system is in the dirty limit. In this drawing, superconducting puddles (in blue) are coupled through a 2DEG (in yellow). The arrows symbolize the local phase of the superconductor}
\label{Figure5}
 \end{figure}

To account for these observations, we propose the following scenario based on the XY model where superconductivity is destroyed by phase fluctuations in a two-dimensional superconductor. We suppose that the system consists in superconducting islands coupled by non-superconducting metallic regions (see Fig.~\ref{Figure5} for a sketch). Indeed, as shown previously\cite{Biscaras:2012eu}, the superconducting O-2DEG is made of two types of carriers, a few Highly Mobile Carriers (HMC) and a majority of Low Mobile Carriers (LMC). We argued that the presence of HMC triggers superconductivity in the system. The density of HMC\cite{Biscaras:2012eu} (in the $10^{12}$~cm$^{-2}$ range) and its evolution with the gate voltage is very similar to the superfluid density directly measured by Bert \emph{et al.}\cite{Bert:2012uv} in \LAO/\STO interfaces. We therefore assume that HMC form the superfluid with intrinsic inhomogeneity due to its very low average density and the associated density fluctuations, and that LMC form the metal which provides long range coupling. The system is therefore described as a disordered array of superconducting puddles of size $L_d$ coupled by a metallic 2DEG. Such a situation has been studied by Spivak, Oreto and Kivelson (SOK)\cite{Spivak:2008cc}. Superconducting phase coherence is governed by the transport properties of the metallic part through the proximity effect, and depends on the 2DEG conductance $G_{2DEG}$. 
 If the coupling is strong enough (high $G_{2DEG}$), puddles can develop full local superconductivity and two phase transitions can be observed corresponding respectively to the puddles themselves and to the disordered array of puddles. In the opposite case (low $G_{2DEG}$), decoupled puddles are always in a fluctuating regime, and only the full array transition can be seen at low temperature. These two regimes can be observed in \LTO/\STO interfaces, and the transition from one to the other can be controlled by the gate voltage. Indeed, the conductance $G_{S}=1/R_{S}$ increases with \VG (inset Fig.~\ref{Figure1}a), and so does the coupling. Region I is therefore the low coupling regime with a single transition ($B_{\times}$=$B_{\mathrm{C}}$), whereas region II refers to the high coupling one with two transitions ($B_{\mathrm{C}}> B_{\times}$). In the later case, $B_{\times}$ is the critical field for the puddle transition and $B_{\mathrm{C}}$ the one for the whole array : as a consequence $T_c$ scales with $B_{\mathrm{C}}$ as observed experimentally.
 
 \indent The SOK model develops a full analysis of the long range coupling between puddles. They introduce the concept of ``optimal puddle''\cite{Spivak:2008cc} to account for the statistical distribution of puddles sizes, and show that the critical magnetic field $B_{\mathrm{C}}$ scales with the coupling parameter $G_{2DEG}$. At low conductance, $B_{\mathrm{C}}\propto G_{2DEG}$, whereas as high conductance, $B_{\mathrm{C}}\propto G_{2DEG}^{1/4}$. To test the model in more details, we plotted the critical field $B_{\mathrm{C}}$ as a function of the conductance $G_S$ on a log-scale in Fig.~\ref{Figure4}c. The data are in good agreement with the theory. Not only the two regimes can be clearly identified, but also the values of the slopes correspond to the calculated one. This is a strong indication that the SOK model is a good representation of the physics involved in these experiments.\\

On this basis, we can now analyse the QPT, remembering that the thermal dephasing length $L_{\Phi}\sim T^{-1/z}$ ($z=1$ here). A cartoon illustrating the situation is shown in Fig.~\ref{Figure5}. Let us first focus on region II as defined in Fig.~\ref{Figure4}. At high temperature, the phase correlation length $L_{\Phi}$ is smaller than the superconducting puddles size $L_d$, and the intra-puddle physics dominates. The critical field $B_{\times}$ corresponds  to the dephasing field $B_d$ on a puddle of size $L_d$ ($B_{d}\sim {\Phi_0}/{L_d^2}$ where $\Phi_0$ is the flux quantum)\cite{Feigelman:472483,Spivak:2008cc}, and does not depend on the microscopic parameters of the system tuned by \VG. This is why it is constant as a function of \VG ($B_{\times}=B_{d}$) as shown in Fig.~\ref{Figure4}a. From the value of $B_d$, we can estimate $L_d$ to be of the order of 100~nm. Since $L_{\Phi}<L_{d}$, the system is in the clean limit, and $z\nu={2/3}<1$ (Fig.~\ref{Figure4}b) as expected from the Harris criteria. 
When lowering the temperature, $L_{\Phi}$ crosses $L_d$ at $T_d$, and the whole array undergoes the transition with a critical field $B_{\mathrm{C}}$. In that case, 
phase fluctuations extend over wide disordered regions and the dirty (2+1)D~XY model applies ($z\nu>1$ as shown in Fig.~\ref{Figure4}b). 
In region I, only the array displays a transition with a single critical field $B_{\mathrm{C}}$. However, as the temperature is lowered, $L_{\Phi}$ also crosses $L_d$, and 
we therefore observe a transition from a clean ($z\nu={2/3}<1$) to a dirty ($z\nu>1$) (2+1)D~XY limit. At this point, it is worthwhile mentioning that, strictly speaking, the transition at $B_{\mathrm{C}}$ is the only true QPT, since the scaling holds down to the lowest temperature. However, at sufficiently high temperature, the transition at $B_{\times}$ displays fluctuations corresponding also to a real QPT which is never reached. Indeed, the diverging coherence length crosses $L_d$, which acts as a cut-off length for the intra-puddle physics. $B_{\times}$ is therefore a crossover field\cite{Spivak:2008cc}, but we consider both transitions as QPT in view of the quantum nature of the fluctuations.

\indent Such an analysis reveals that the tunable \LTO/\STO epitaxial interface appears as a unique system to study superconducting QPT in 2D, which is well described by the (2+1)D~XY model as expected. The role of disorder on the critical exponents is clearly evidenced, together with the possibility of observing multiple QPT as previously proposed theoretically \cite{Spivak:2008cc,Spivak:2001bk,Feigelman:472483}. These models have been developed in the framework of the so-called ``fermionic scenario''\cite{Finkelshtein:1987vo,Yazdani:1995ut}, where superconductivity disappears because Cooper pairs are destroyed, as opposed to the ``bosonic scenario''\cite{Fisher:1990vl}, where they localize and form an insulator. In the later case, the QPT takes place for critical resistances $R_{\mathrm{C}}$ close to the quantum resistance $R_Q=h/4e^2 \simeq6.5$~k$\Omega/\square$, whereas in the former one it can occurs at much lower resistances\cite{Spivak:2008cc,Spivak:2001bk,Feigelman:472483}. Steiner \emph{et al.}\cite{Steiner:2008ke} compiled experimental results in the literature and clearly evidenced the two behaviours : the QPT separates a superconducting phase from a weakly localizing metal in the less disordered systems ($R_{\mathrm{C}}\leq R_Q$), and from an insulator in strongly disordered materials ($R_{\mathrm{C}}\sim R_Q$). Our data ($0.04\leq ({R_{\mathrm{C}}}/{R_Q}) \leq0.4$) fully agree with this picture, and the non-superconducting phase is indeed always a weakly localizing metal.\\

As a conclusion, we  showed that the superconducting O-2DEG at the \LTO/\STO interface undergoes a QPT from a superconductor to a weakly localizing metal (fermionic scenario) upon applying a perpendicular magnetic field, driven by phase fluctuations and well described by the (2+1)D~XY model, as expected. By tuning the gate voltage, it is possible to explore the ``clean'' and ``dirty'' regimes according to the Harris criteria, with a critical exponents product $z\nu={2/3}$ in the former case, in agreement with previous evaluations\cite{Caviglia:2008uh}\cite{Schneider:2009gt}, and greater than one otherwise. The system is well described by a disordered array of superconducting puddles coupled by a 2DEG, which can exhibit two QPT, one related to ``regional'' or local ordering, and another one corresponding to long range phase coherence, as proposed theoretically\cite{Spivak:2008cc,Spivak:2001bk,Feigelman:472483}. The key parameter, that is the coupling constant (the 2DEG conductance), can be tuned at will to explore the phase diagram of the system. This is important in the context of recent studies of strongly disordered 2D superconductors where intrinsic inhomogeneities appear at mesoscopic scales\cite{Sacepe:2011jm,Bouadim:2011hx,Spivak:2008cc}, with coexistence of superconducting and non-superconducting regions. Recent work on artificial ordered metallic networks addresses this issue\cite{Eley:2011by}, but does not reach the insulating state. Our study opens the way for exploring the physics of disordered superconductors, and beyond, the more general problem of phase coherence in multiscale systems such as strongly correlated materials which are phase separated\cite{Dagotto:2005ip} or spin-textured\cite{Ruben:2010bx}.\\

\indent
The Authors gratefully thank C. di Castro, C. Castellani, L. Benfatto and C. Kikuchi-Marrache for stimulating discussions. This work has been supported by the R\'egion Ile-de-France in the framework of CNano IdF and Sesame programs. Part of this work has been supported by Euromagnet II. Research in India was funded by the Department of Information Technology, Government of India.\\

\newpage

\end{document}